\documentstyle[eqsecnum,preprint,aps,epsfig]{revtex}
\tightenlines

\begin{document}
\draft

\title{Localization of Kaluza-Klein gauge fields on a brane.}

\author{Andrey Neronov 
\footnote{e-mail:neronov@theorie.physik.uni-muenchen.de}}

\address{Theoretische Physik, Universit\"at Munchen, 80333, Munich, Germany}

\date{\today}

\maketitle


\begin{abstract}
In phenomenological models with extra dimensions there is a natural symmetry 
group associated to a brane universe, -- the group of rotations of normal 
bundle of the brane. We consider Kaluza-Klein gauge fields corresponding to 
this group and show that they can be localized on the brane in models with 
warped extra dimensions. These gauge fields are coupled to matter fields 
which have nonzero rotation moment around the brane. 
In a particular example of a four-dimensional 
brane embedded into  six-dimensional asymptotically anti-deSitter space, 
we calculate effective four-dimensional coupling constant between the 
localized fermion zero modes and the Kaluza-Klein gauge field.  
\end{abstract}

\narrowtext

\newpage

\section{Introduction.}

A possibility to treat observable gauge fields as arising from dimensional 
reduction of
higher-dimensional General Relativity has a long history \cite{appelquist}.
If we suppose that the space-time is a direct product of 
four-dimensional Minkowsky space $R^4$ with coordinates $x^\mu, \ \mu=0,..,3$ 
on an $n$-dimensional 
``internal space'' $K^n$ parameterized by coordinates $y^a,\  a=1,..,n$, 
the four-dimensional gauge fields $A_\mu^a$ can be described by the $\mu a$
components of the higher-dimensional metric
\begin{equation}
\label{kk}
g_{AB}=\left(
\begin{array}{cl}
g_{\mu\nu}+g_{ab}A^a_\mu A^b_\nu & \vline \ g_{ab}A^b_\mu\\
\hline
g_{ab}A^b_\mu &\vline\  g_{ab}
\end{array}
\right)
\end{equation}
In this simple form such an approach to the gauge interactions faces 
certain problems as,  for example, the problem of obtaining  
realistic pattern of four-dimensional fermions with different charges with 
respect to $A_\mu^a$  
after dimensional reduction of higher-dimensional fermion 
fields \cite{witten}. The problem of chiral 
fermions can be solved in models with  additional 
fundamental higher-dimensional gauge fields fields in 
topologically nontrivial configurations \cite{salam},
or in models with  noncompact internal spaces \cite{wetterich}.

If we deal with noncompact internal spaces, we face a problem how to explain 
the observable four-dimensional structure of the universe. One is forced 
to consider the observable space-time as a surface (brane) 
embedded in a higher-dimensional manifold \cite{rubakov,gibbons}. 
The idea of the  ``brane world'' received a considerable attention recently
due to new developments in the string theory. A new 
impulse to higher-dimensional model building was given by papers 
\cite{dimopoulos,randall} 
where the possibility of having ``large'' or 
infinite extra dimensions was considered in relation to the hierarchy problem 
of particle physics. The observational consequences of the brane world 
picture for the accelerator physics \cite{hep}, astrophysics \cite{astro}
and cosmology \cite{cosmology,ivo} 
were extensively studied recently. The possibility of solving the fermion 
mass hierarchy \cite{fermion_hierarchy}, cosmological constant \cite{constant},
supersymmetry breaking \cite{supersymmetry} problems with large extra 
dimensions was also discussed.

An important problem which must be addressed in phenomenological models with
a brane universe is how the observable matter fields of the Standard Model of 
particle physics are localized on the brane. The mechanism of localization of
fermions is known since a long time \cite{jackiw} while mechanisms of 
localization of gauge fields \cite{dvali-shifman,dubovsky,oda,gababadze} 
were found recently.  

In this paper we discuss the question whether it is possible to localize the 
Kaluza-Klein fields $A_\mu^a$ on the brane and to treat them as observable 
gauge gauge fields. If the brane is embedded in a space-time with $n\ge 2$
extra dimensions there is  
a ``natural'' symmetry group associated to the brane: 
the group of rotations of its normal 
bundle. In the case of $n=2$ extra dimensions the brane can be treated as a 
string-like defect in a higher-dimensional space-time.
The symmetry group 
of the normal bundle is $U(1)$ and corresponds to  
rotations of normal vectors around the brane.
We analyze the gauge field $A_\mu$ which corresponds to this
symmetry in Kaluza-Klein Ansatz (\ref{kk}). We show that $A_\mu$ can be 
localized on a four-dimensional string even if the extra dimensions are 
noncompact and have infinite volume, as it is, for example, in the 
space-times considered in \cite{gibbons}. We consider coupling of 
the field $A_\mu$ to the localized fermion fields  in 
a   particular model where a four-dimensional brane is embedded into 
a six-dimensional 
asymptotically anti-deSitter space-time \cite{gregory,shaposhnikov}. 
In this space-time both the Kaluza-Klein field $A_\mu$ which corresponds to 
the $U(1)$ symmetry of rotations of the normal bundle and 
the fermion zero modes which have nonzero momentum of 
rotation around the string are localized. We calculate the  
effective four-dimensional fine structure constant $\alpha$ in this model and 
find that it is related to the curvature radius of anti-deSitter space-time
and to the thickness of the brane.

\section{Gauge fields associated to the group of rotations of normal bundle.}

In the models with large or infinite extra dimensions the 
gauge fields $A_\mu^a$ which appear by dimensional reduction mechanism  
(\ref{kk}) are not considered as observable gauge fields of the Standard Model 
because of  the following reasons. The Kaluza-Klein fields 
$A_\mu^a$ (\ref{kk}) are, in general, not confined to the brane and 
can propagate in the bulk. 
The tower of Kaluza-Klein excitations 
of these fields starts from very low masses. For example, in 
the model with submillimeter extra dimensions \cite{dimopoulos} the mass 
scale of Kaluza-Klein excitations is of order of $10^{-3}$ eV which means that
there must be corrections to Coulomb law at distances of order of $1$ mm. 
In the case of noncompact extra 
dimensions one encounters another difficulty. Consider the case of 
single extra dimension. A four-dimensional brane $M^4$ is embedded in 
higher dimensional bulk ${\cal M}^5$ as a level surface of some function
$F(x^A),\ A=0,..,5$
\begin{equation}
M^4: \{ F(x^A)=0\}
\end{equation}
If we take coordinates $x^\mu$ on the brane 
and the function $F$ as a coordinate system  in the vicinity
of the brane, the metric on ${\cal M}^5$ takes the form
\begin{equation}
\label{nc}
ds^2=g_{\mu\nu}dx^\mu dx^\nu+g_{44}dF^2
\end{equation}
since vector $\vec N_A=F_{,A}$ is a normal to the surfaces $F(x^A)=const$.
Comparing (\ref{nc}) with (\ref{kk}) we find that it is always possible 
to find a coordinate system in the neighborhood of the brane in which
\begin{equation}
A_\mu(x^A)\equiv 0
\end{equation}
If the extra dimension is noncompact,
 such a coordinate system can be chosen 
globally and $A_\mu$ can be always removed by a gauge 
transformation of five-dimensional theory.

The situation changes if the number of extra dimensions is $n\ge 2$. In this 
case we have $4+n$ coordinate transformations
\begin{equation}
x^A\rightarrow x^A+\xi^A(x)
\end{equation}
at our disposal. Therefore we can impose $4+n$ gauge conditions on the metric. 
For example, $g_{4\mu}=0$, $g_{44}=1$, $g_{4a}=0, a=1,..,(n-1)$. After such a 
gauge fixing the metric becomes
\begin{equation}
\label{mod}
g_{AB}=\left(
\begin{array}{ccc}
g_{\mu\nu}+g_{ab}A^a_\mu A^b_\nu &\vline\  0\ \vline& g_{ab}A^b_\mu\\
\hline
0&\vline \ 1\ \vline&0\\
\hline
g_{ab}A^b_\mu &\vline \ 0 \ \vline& g_{ab}
\end{array}
\right)
\end{equation} 
The fields $A_\mu^a, a=1,.., (n-1)$ can not be removed by a coordinate transformation.
These fields have a clear geometrical meaning. Let the coordinate 
$r=x^4$ be a distance from the brane $M^4$ placed at $r=0$. Then the fields 
$A_\mu^a$ are the gauge fields associated to the symmetry group $G$ 
of rotations of the normal bundle of the brane. 
In the simplest case of a thin four-dimensional brane $M^4$
embedded into a higher-dimensional manifold ${\cal M}^{4+n}$ the coordinates 
$\theta^a, a=1,.., (n-1)$ parameterize a small $(n-1)$-dimensional 
sphere around the point 
$r=0$ of location of the brane and the group of the normal bundle
is $SO(n)$.  The group $G$ may be more involved if we consider
branes of nonzero thickness.  The stress-energy 
tensor of the brane is different from zero in a region  
$M^4\times K^n=\{r\le\epsilon\}$ for some small $\epsilon$. 
The surface $r=\epsilon$ is a $3+n$-dimensional boundary 
$M^4\times \partial K^n$ of the thick brane.
The group $G$ is the group of isometries 
of $\partial K^n$. It can be quite arbitrary. For example, in the model 
considered in \cite{constant}, this group is $SU(2)\times U(1)$. 

If we 
introduce a coordinate system $(x^\mu, \theta^a)$ on 
$M^4\times \partial K^n$,
the background metric (with $A_\mu^a=0$) in the vicinity of the brane 
can be written in the form 
\begin{equation}
\label{metric}
ds^2=e^{\nu(r)}\eta_{\mu\nu}dx^\mu dx^\nu+e^{\lambda(r)} dr^2+
\varphi(r)g_{ab}(\theta)d\theta^a d\theta^b
\end{equation}
where $g_{ab}(\theta)$ is a $G$-symmetric metric on $\partial K^n$. 
The functions $\nu(r),\lambda(r)$ and $\varphi(r)$ are found from the 
$4+n$-dimensional Einstein equations
\begin{equation}
\label{einstein}
R_{AB}-\frac{1}{2}g_{AB}R+\Lambda g_{AB}=8\pi G_{4+n}T_{AB}
\end{equation}
where $G_{4+n}, \Lambda$  are $4+n$-dimensional gravitational and cosmological 
constants and 
$T_{AB}$ is stress-energy tensor generated by the brane.

We are interested in dynamics of small perturbations $A_\mu^a$ 
around the background metric $g_{AB}^{bg}$ (\ref{metric}).  In what follows we 
will analyze in detail the case of two extra dimensions when there is 
single angular coordinate $\theta$. The generalization on the case $n>2$ is 
straightforward. 

\section{Equations of motion for $A_\mu$.}

The Ricci tensor for the metric (\ref{metric}) perturbed by the fields 
$A_\mu$ (\ref{mod}) in the linear in $A_\mu$ approximation has the form  
\begin{eqnarray}
R_{\mu\nu} &=& -e^{\nu-\lambda}\eta_{\mu\nu}\left(\frac{\nu''}{2}
-\frac{\nu'\lambda'}{4}+{\nu'}^2+\frac{\nu'\varphi'}{4\varphi}\right)+
\frac{1}{2}(\dot A_{\mu,\nu}+\dot A_{\nu,\mu})
\\
R_{\mu \theta} &=& \frac{e^{-\nu}\varphi}{2}(A_{\nu,\mu\nu}-A_{\nu,\mu\mu})+
\frac{\varphi e^{-\lambda}}{2}\left[-A_\mu''+
\left(\frac{\lambda'}{2}-\nu'-
\frac{5\varphi'}{2}\right)A_{\mu}'+\right.\nonumber\\
&&\left.\left(\frac{
{\varphi'}^2}{2\varphi^2}
-\frac{\varphi''}{\varphi}+\frac{\varphi'\lambda'}{2\varphi}-
\frac{\varphi'\nu'}{\varphi}\right)A_{\mu}
\right]
\\
R_{\mu r} &=& \frac{1}{2\varphi} (\varphi \dot A_\mu)'
-\frac{\nu'}{2}\dot A_\mu 
\\
R_{rr} &=& -2\nu''-{\nu'}^2+\nu'\lambda'
+\frac{1}{2}\left(
\frac{{\varphi'}^2}{2\varphi^2}+\frac{\varphi'\lambda'}{2\varphi}
-\frac{\varphi''}{\varphi}
\right)
\\
R_{r\theta} &=& \frac{\varphi e^{-\nu}}{2}A'_{\mu,\mu}
\\
R_{\theta\theta} &=& \varphi e^{-\nu}
\dot A_{\mu,\mu}+e^{-\lambda}\left(
\frac{\varphi'\lambda'}{4}
-\frac{{\varphi'}^2}{4\varphi}-\frac{\varphi''}{2}
-\varphi'\nu'\right) 
\\
\end{eqnarray}
When $A_\mu=0$ we get from (\ref{einstein}), the expression of 
$\nu(r), \lambda(r)$ and $\varphi(r)$ through the brane stress-energy 
tensor $T_{AB}$.
In the presence of  $A_\mu$ the $\mu\nu$, $\mu r$, $r\theta$ and 
$\theta\theta$ components of Einstein equations (\ref{einstein}) become
\begin{eqnarray}
\label{one}
\dot A_{(\mu,\nu)}&=&0\\
\label{two}
\left(e^{-\nu}\varphi\dot A_\mu\right)'&=&0\\
\label{three}
\dot A_{\mu,\mu}=A_{\mu,\mu}'&=&0
\end{eqnarray}
Here prime denotes the derivative with respect to $r$ and dot denotes the 
$\theta$ derivative.
The only nontrivial equation is the 
$\mu\theta$ component of Einstein equations. In the most simple case of flat 
extra dimensions when the background metric is
\begin{equation}
\label{flat}
ds^2=\eta_{\mu\nu}dx^\mu dx^\nu+dr^2+r^2 d\theta^2
\end{equation}
it  reduces to
\begin{equation}
\label{four}
A_\mu''+\frac{5}{r}A_\mu'+F_{\mu\nu,\nu}=0
\end{equation}   
where we have denoted 
\begin{equation}
F_{\mu\nu}=A_{\mu,\nu}-A_{\nu,\mu}
\end{equation}
First of all we can see that the ``zero mode''  
$A_\mu(x^\nu)$ which does not depend on 
$r,\theta$ and satisfies
\begin{equation}
\label{maxwell}
F_{\mu\nu,\nu}=0
\end{equation}
is a solution of the six-dimensional Einstein equations. The last system 
of equations is just the four-dimensional Maxwell equations
on electromagnetic field $A_\mu(x)$.  
The $U(1)$ gauge group of electromagnetism 
is identified with the group $SO(2)$ of rotations of the normal 
bundle of the brane. Indeed, let us  make a coordinate transformation
\begin{equation}
\label{rotation}
\tilde \theta=\theta+v(x^\mu)
\end{equation}
which rotates normal vectors to the brane at the point $x^\mu$ on a 
small angle $v(x^\mu)$. The background metric (\ref{flat}) 
in new coordinates takes the form
\begin{eqnarray}
\label{flat1}
ds^2=(\eta_{\mu\nu}+r^2v_{,\mu}v_{,\nu})dx^\mu dx^\nu+
dr^2
+r^2d\tilde\theta^2-2r^2v_{,\mu}dx^\mu d\tilde\theta
\end{eqnarray}
from where we see that in new coordinates
\begin{equation}
\label{pure}
A_\mu=-v_{,\mu}
\end{equation}   
Since the metric (\ref{flat1}) differs from (\ref{flat}) on coordinate 
transformation, it is again a solution of the Einstein equations. 

Obviously, by the same way of reasoning, the pure gauge configuration
(\ref{pure}) must be a solution of the perturbed Einstein equations 
on a general background of the form (\ref{metric}). The $\mu\theta$ component
of the Einstein equations in the general case is
\begin{eqnarray}
\label{general}
\left(e^{\nu-\lambda/ 2}\varphi^{5/2} A_\mu'\right)' 
+e^{\lambda/2}\varphi^{5/2}F_{\mu\nu,\nu}= 0  
\end{eqnarray} 
Taking the $\theta$ derivative of (\ref{general}) and combining it with
(\ref{two}) we find that 
$A_\mu$ does not depend on $\theta$. Making a coordinate transformation 
(\ref{rotation}) we can  always fix the coordinate system in such a way that 
 the condition 
\begin{equation}
\label{gauss}
A_{\mu,\mu}=0
\end{equation}
is satisfied on a particular surface $r=r_0$. 
Then, from (\ref{three}) we conclude that 
the condition (\ref{gauss}) remains valid for all $r$. The equation 
(\ref{general}) in the gauge (\ref{gauss})  reduces after
the Fourier transform $\tilde A_\mu(p, r)=\int d^4x e^{-ip_\nu x^\nu}A_\mu(x, r)$ to
\begin{equation}
\label{fourier}
\left(e^{\nu-\lambda/2}\varphi^{5/2}\tilde A_\mu'\right)'+
e^{\lambda/2}\varphi^{5/2}
m^2\tilde A_\mu=0
\end{equation}
where we have denoted $m^2=-p_\mu p^\mu$. This equation can be  
solved, for example, in the case of flat extra dimensions (\ref{flat}), 
$\lambda=\nu=0$, $\varphi=r^2$.
The modes with $m^2\not= 0$ which are regular at $r=0$ have the form 
\begin{equation}
\label{massive}
A_\mu=\frac{C}{r^2} J_2(mr)
\end{equation}
where $J_2(z)$ is the Bessel function and $C$ is an arbitrary constant.

Although the zero mode $A_\mu(x)$ is always a solution of Einstein equations
locally, it can fail to be a global solution if the rotation invariance 
(\ref{rotation}) is spontaneously broken. A simple example of this is 
when  the flat extra dimensions (\ref{flat}) are compactified on 
two-dimensional torus $T^2$.

\section{Localization of $A_\mu$ on the brane.}

If the rotation symmetry (\ref{rotation}) is not broken, the zero mode of the 
field $A_\mu$ is a global solution of higher-dimensional Einstein equations 
which, from the four-dimensional point of view, describes massless gauge field 
propagating along the brane. The equation (\ref{fourier}) is  a 
Strum-Liouville equation
and the  zero mode $A_\mu(x^\nu)$ is its normalizable solution
in a space-time (\ref{metric})
which satisfies the condition
\begin{equation}
\label{integral}
N=\int dr e^{\lambda/2}\varphi^{5/2}<\infty
\end{equation}
In such a space-time the massless field $A_\mu(x^\nu)$ mediates a Coulomb-like
interaction between particles localized on the brane. For example, if  
we take a space-time with compact extra dimensions with topology of a sphere 
$S^2$, the bulk metric is 
\begin{equation}
ds^2=\eta_{\mu\nu}dx^\mu dx^\nu+\frac{1}{(1+r^2/a^2)^2}dr^2+
\frac{r^2}{(1+r^2/a^2)^2}d\theta^2
\end{equation}
where $a$ is the radius of the sphere. This corresponds to
\begin{eqnarray}
\nu&=&0\nonumber\\
\lambda&=&-2\log(1+r^2/a^2)\\
\varphi&=& r^2/(1+r^2/a^2)\nonumber
\end{eqnarray}  
and the integral $N$ (\ref{integral}), obviously, converges.

Zero mode of the field $A_\mu$ can be normalizable also in space-times with 
noncompact extra dimensions. Let us take, for example, a space-time considered 
in \cite{gibbons}.  This space-time  is a solution of six-dimensional 
Einstein equations  coupled to cylindrically symmetric magnetic field
\begin{equation}
\label{em}
f_{r\theta}=B_0\sqrt{\varphi} e^{\lambda/2-2\nu}
\end{equation}
where $B_0$ is an arbitrary magnetic field strength.
In the case when the cosmological constant in the bulk and on the brane are 
zero, the space-time metric is 
\begin{equation}
\label{gibbons}
ds^2=\left(1+\frac{r^2}{a^2}\right)^{2/3}\left(
\eta_{\mu\nu}dx^\mu dx^\nu+dr^2\right)+
\frac{r^2d\theta^2}{(1+r^2/a^2)^2}
\end{equation}
where the parameter $a$ is related to the field strength (\ref{em})
\begin{equation}
a^2=\frac{2}{3\pi G_6 B_0^2}
\end{equation}
The geometry of extra dimensions is schematically presented on Fig.1. The size
$R_\theta$
of the circle parameterized by $\theta$ shrinks to zero as 
$r\rightarrow\infty$.

\vspace{5mm}
\centerline{\epsfig{file=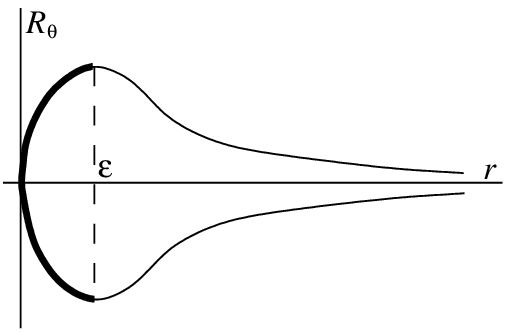,width=7cm}}
{\footnotesize\textbf{Figure 1:} The geometry extra dimensions. }
\vspace{5mm}

The integral $N$ (\ref{integral}) converges
\begin{equation}
N=\int\limits_0^\infty\frac{r^5dr}{(1+r^2/a^2)^{14/3}}=\frac{27 a^6}{440}
\end{equation}  
and the zero mode $A_\mu(x^\nu)$ is a normalizable solution of (\ref{fourier})
Note, that the volume of extra dimensions 
\begin{equation}
V=\int\sqrt{-g}dr d\theta=\int \left(1+\frac{r^2}{a^2}\right)^{2/3}rdrd\theta
\end{equation}
is infinite so that the metric (\ref{gibbons})
does not localize gravity. 

It is interesting
to consider situation when both gravity and the gauge field are localized 
on the brane. This can be achieved in a six-dimensional generalization 
of Randall-Sundrum model which was considered in 
\cite{gregory,shaposhnikov,oda}. The 
space-time is a solution of Einstein equations (\ref{einstein}) with negative
cosmological constant $\Lambda$. A string of finite size $\epsilon$ 
is placed at $r=0$ so that the stress-energy tensor $T_{AB}$ is different 
from zero in a small region $0\le r<\epsilon$ around $r=0$. Outside 
the string in the region $\epsilon<r<\infty$ the space-time metric is 
the six-dimensional anti-deSitter metric 
\begin{equation}
\label{ads}
ds^2=e^{-cr}\eta_{\mu\nu}dx^\mu dx^\nu+dr^2+e^{- cr}R^2d\theta^2
\end{equation} 
Here $R$ is an arbitrary constant which sets the size of the circle $S^1$ 
parameterized by $\theta$.
Parameter $c$ is related to the bulk cosmological constant (\ref{einstein}) as
\begin{equation}
c=\sqrt{-\frac{2\Lambda}{5}}
\end{equation}
The geometry of extra dimensions in this case is similar to that of space-time
(\ref{gibbons}): the size of the circle $S^1$ parameterized by $\theta$ 
goes to zero as $r\rightarrow\infty$ (see Fig.1). 
The metric inside the brane is determined by the detailed structure of 
the brane stress-energy tensor $T_{AB}(r)$.

The equation (\ref{fourier}) on $A_\mu$ reduces for the metric (\ref{ads}) to 
\begin{equation}
\label{general1}
\left(e^{-7cr/2}\tilde A_\mu'\right)'+e^{-5cr/2}m^2 \tilde A_\mu=0
\end{equation}
Its general solution in the gauge (\ref{gauss}) is
\begin{eqnarray}
\label{hom}
\tilde A^0_\mu= e^{7r/4}  
\left(C_1J_{7/2}\left(\frac{2m}{c}e^{cr/2}\right)+
C_2Y_{7/2}\left(\frac{2m}{c}e^{cr/2}\right)\right)
\end{eqnarray}

The zero mode solution $A_\mu(x)$ of (\ref{general1}) is a normalizable
solution of (\ref{general1}) 
since the integral $N$ (\ref{integral}) is converging
\begin{equation}
\label{nor}
N=\frac{2R^5}{5c}
\end{equation}
Note that the zero mode of $A_\mu$ is normalizable even in a more general 
space-time of ``global string'', considered in \cite{oda,shaposhnikov1}
\begin{equation}
\label{global}
ds^2=e^{-c_1r}\eta_{\mu\nu}dx^\mu dx^\nu+dr^2+e^{-c_2 r}d\theta^2
\end{equation}
Here $c_1$ and $c_2$ are arbitrary positive constants related to the 
stress-energy tensor $T_{AB}(r)$ of the string which is nonzero even when 
$r\rightarrow\infty$.
 
\section{Matter fields bound  to the brane.}

The field $A_\mu$ interacts with matter fields bound to the brane. As an 
example we consider fermion fields 
propagating on the background (\ref{metric}). 

Consider a six-dimensional Dirac spinor $\Psi$ which satisfies the equation
\begin{equation}
\label{dirac}
\Gamma^A D_A\Psi=0
\end{equation}
The six-dimensional gamma matrixes $\Gamma^A$ are defined 
with the help of vielbein $E^A_{\hat B}$ and flat space gamma matrices 
$\Gamma^{\hat A}$
\begin{equation}
\Gamma^A=E^A_{\hat B}\Gamma^{\hat B}
\end{equation}
(the indexes with hat are six-dimensional Lorenz indexes).
 The covariant derivative is defined as 
\begin{equation}
D_A\Psi=\Psi_{,A}+\frac{1}{2}\omega^{\hat A\hat B}_A\sigma_{\hat A\hat B}\Psi
\end{equation}
where $\omega_A^{\hat B\hat C}$ is the spin connection expressed through
vielbein $E^A_{\hat B}$ and
$\sigma_{\hat B\hat C}=\frac{1}{4}\left[\Gamma_{\hat B}\Gamma_{\hat C}\right]$.
Taking the coordinate vielbein for the metric (\ref{metric})
\begin{eqnarray}
\label{viel}
E_\mu^{\hat \alpha}&=&e^{\nu/2}\delta^{\hat \alpha}_\mu\nonumber\\
E^{\hat r}_r&=&e^{\lambda/2}\\
E^{\hat \theta}_{\theta}&=&\sqrt{\varphi}\nonumber
\end{eqnarray}
we find
\begin{eqnarray}
\omega_\mu^{\hat A\hat B}\sigma_{\hat A\hat B}&=&
\frac{\nu'}{2}e^{(\nu-\lambda)/2}\Gamma_{\hat \mu}\Gamma_{\hat r}
\nonumber\\
\omega_\theta^{\hat A\hat B}\sigma_{\hat A\hat B}
&=&\frac{\varphi'}{2\sqrt{\varphi}}e^{-\lambda/2}
\Gamma_{\hat \theta}\Gamma_{\hat r}
\end{eqnarray}

Let us consider the solutions of (\ref{dirac}) which satisfy the condition
\begin{equation}
\label{chir6}
\Gamma_{\hat 0}...\Gamma_{\hat 3}\Gamma_{\hat r}\Gamma_{\hat\theta}\Psi=\Psi
\end{equation} 
and can be presented in the form
\begin{equation}
\label{psi}
\Psi=e^{iq\theta}\chi(r)\psi(x^\mu)
\end{equation}
where $q$ is an integer. The two-component spinor  $\chi(r)$  
satisfies 
\begin{equation}
\label{chir2}
i\Gamma_{\hat r}\Gamma_{\hat \theta}\chi=-\chi
\end{equation}
while the four-component spinor $\psi(x)$ is chiral in four-dimensional sense
\begin{equation}
\label{chir4}
i\Gamma_{\hat 0}...\Gamma_{\hat 3}\psi=\psi
\end{equation}
and satisfies 
the four-dimensional massless Dirac equation
\begin{equation}
\label{massless}
\Gamma^{\hat \mu}\partial_\mu\psi=0
\end{equation}
From (\ref{dirac}) we derive 
an equation on $\chi$
\begin{equation}
\label{chi}
\chi'+\left(\nu'+\frac{\varphi'}{4\varphi}
+\frac{e^{\lambda/2}}{\sqrt{\varphi}}q\right)
\chi=0
\end{equation}
which is readily integrated 
\begin{equation}
\label{sss}
\chi(r)=const\cdot\varphi^{-1/4} \exp\left\{-\nu-q
\int\limits_{r_0}^r dr'\frac{e^{\lambda/2}}{\sqrt{\varphi}}\right\}
\end{equation}
The solutions of (\ref{dirac}) must be normalizable with respect to the
norm
\begin{equation}
\label{norm}
\left<\Psi,\Psi\right>=
\int d^4x dr d\theta\sqrt{-g}\Psi^*\Psi=
\int dr e^{2\nu+\lambda/2}\sqrt{\varphi}|\chi|^2
\end{equation}

In the case of  space-time (\ref{ads}) the normalized solutions of equation
(\ref{chi}) are given by 
\begin{equation}
\label{rad}
\chi_q(r)=\frac{4 q^{3/2} }{R\sqrt{8q^2+4qcR+c^2R^2}}
 e^{5cr/4}\exp\left\{\frac{2q}{cR}(1-e^{cr/2})
\right\}
\end{equation} 
The solution with $q=0$ with trivial dependence on $\theta$ grows at 
large $r$ and is not localized on the brane. But the solutions with 
$q>0$ decrease far from the brane.  Here we have neglected a contribution 
from the region  $0\le r<\epsilon$ inside the brane into the 
integral (\ref{norm}). In principle, the requirement of convergence
of (\ref{norm}) in the limit $r\rightarrow 0$ will impose restrictions on 
the number of normalizable zero mode solutions of (\ref{dirac}).

\section{Coupling of $A_\mu$ to matter fields.}

The fermion field $\Psi$ (\ref{dirac}) produces a stress-energy tensor 
\begin{equation}  
\label{stt}
T^{\Psi}_{AB}=\frac{i}{2}\left(
\overline \Psi\Gamma_{(A} D_{B)}\Psi-D_{(A}\overline\Psi
\Gamma_{B)}\Psi\right)
\end{equation} 
The $\mu\theta$ component of $T_{AB}^{\Psi}$ is
\begin{eqnarray}
\label{muth}
T_{\mu\theta}^\Psi&=&\frac{i}{4}e^{\nu/2}\left(\overline\Psi\Gamma_{\hat \mu}
\Psi_{,\theta}-\overline\Psi_{,\theta}\Gamma_{\hat \mu}\Psi\right)+
\frac{i\sqrt{\varphi}}{4}\left(\overline\Psi\Gamma_{\hat \theta}
\Psi_{,\mu}-\overline\Psi_{,\mu}\Gamma_{\hat \theta}\Psi\right)+\nonumber\\&&
\frac{i\sqrt{\varphi}}{4}e^{(\nu-\lambda)/2}
\left(\nu'-\frac{\varphi'}{\varphi}\right)\overline\Psi\Gamma_{\hat \mu}
\Gamma_{\hat r}\Gamma_{\hat\theta}\Psi
\end{eqnarray}
If we restrict our attention to the $\Psi$ configurations of the form
(\ref{psi}), (\ref{chir2}), (\ref{chir4}), then (\ref{muth}) reduces to
\begin{equation}
\label{str}
T_{\mu\theta}^\Psi=\frac{e^{\nu/2}|\chi(r)|^2}{2}
\left(m-e^{-\lambda/2}
\frac{\sqrt{\varphi}}{2}\left(\nu'-\frac{\varphi'}{\varphi}\right)\right)
\overline \psi\Gamma_{\hat \mu}\psi
\end{equation}
We 
see that $T_{\mu\theta}^\Psi$ is proportional to the four-dimensional current
\begin{equation}
\label{jmu}
j_\mu=\overline\psi\Gamma_{\hat \mu}\psi
\end{equation}
Substituting the stress-energy tensor (\ref{str}) into the Einstein equations
(\ref{einstein}) we find that equation (\ref{fourier}) on $A_\mu$ is modified
\begin{eqnarray}
\label{fourier1}
\left(e^{\nu-\lambda/2}\varphi^{5/2}\tilde A_\mu'\right)'+
e^{\lambda/2}\varphi^{5/2}
m^2\tilde A_\mu=
-8\pi G_6e^{2\nu+\lambda/2}\varphi^{3/2}|\chi|^2
Q\tilde j_\mu
\end{eqnarray}
where $\tilde j_\mu$ is the Fourier transform 
$\tilde j_\mu(p, r)=\int d^4p e^{-ip_\mu x^\mu} j_\mu(x, r)$ of the 
four-dimensional current. We also have denoted
\begin{equation}
Q=q-e^{-\lambda/2}\frac{\sqrt{\varphi}}{2}\left(\nu'-
\frac{\varphi'}{\varphi}\right)
\end{equation}
We can write a solution of this equation in terms of a Green function
$\tilde \Delta(p, r, r')$
\begin{equation}
\label{ammm}
\tilde A_\mu=-8\pi G_6\tilde j_\mu
\int dr' \tilde \Delta (p, r, r')
Qe^{2\nu+\lambda/2}\varphi^{3/2}|\chi|^2 
\end{equation}
The Green function of equation (\ref{fourier1}) 
is expressed in terms of normalized 
solutions $A^0_m(r)$ of homogeneous
equation (\ref{fourier}) for a fixed $m^2$
\begin{equation}
\tilde \Delta(p, r, r')=\frac{1}{Np^2}+
\sum_{m^2>0}\frac{A^0_m(r)A^0_m(r')}{p^2+m^2}
\end{equation}
where $N$ is the normalization constant (\ref{integral}). If we are interested 
in the behavior of $A_\mu$ at large distances from the source, we can 
restrict attention to the limit $p^2\rightarrow 0$ in which zero mode solution
of (\ref{fourier}) gives 
the leading contribution into the Green function $\tilde \Delta$. In 
this limit
\begin{equation}
\label{soll}
\tilde A_\mu \approx\frac{8\pi G_6}{N}\frac{\tilde j_\mu(p)}{p^2}
\int dr' Q\varphi^{3/2}e^{2\nu+\lambda/2}|\chi(r')|^2
\end{equation} 
In the case of asymptotically anti-deSitter space-time (\ref{ads}) 
 the normalization constant 
$N$ is given by (\ref{nor}) while the charged zero modes of Dirac field
have the radial profile $\chi(r)$ given by (\ref{rad}).  In the limit 
\begin{equation}
\label{conn}
cR\ll 1
\end{equation}
when the curvature radius $\rho=c^{-1}$ 
of anti-deSitter space is much larger than the thickness of the brane $R$,
the expression (\ref{soll}) for $\tilde A_\mu$ reduces to
\begin{equation}
\label{amm}
\tilde A_\mu=\frac{20\pi cG_6}{R^3}\frac{ (q\tilde j_\mu)}{p^2}
\end{equation}  
Comparing the last equation to the standard expression of the four-dimensional 
Maxwell theory we find the effective fine structure constant
\begin{equation}
\label{al}
\alpha=\frac{5 c G_6}{R^3}
\end{equation}

 From (\ref{rad}) we can see that the fermion 
zero modes with nonzero charge $q$ are localized in a region with a size of 
order of the brane thickness $R$. If $R$ is small enough we can approximate 
the profile $\chi(r)$ by delta function 
\begin{equation}
|\chi|^2\approx \frac{1}{R}\delta(r)
\end{equation}
If we are interested in interactions of particles localized on the brane   
we need only the 
expression for the Green function $\tilde \Delta$ at $r, r'=0$. The exact 
expression for $\tilde\Delta$ can be found in a way similar 
to the one used in \cite{giddings} 
for the calculation of the Green function for gravitational perturbations 
of Randall-Sundrum model. For $r, r'=0$ we get
\begin{equation}
\tilde \Delta (p)=\frac{5c}{2k^2}-\frac{1}{k}\frac{H_{3/2}^{(1)}
(2k/c)}{
H_{5/2}^{(1)}(2k/c)}
\end{equation}
where $k^2=-p^2$.
Expressing the Hankel functions $H_{3/2}^{(1)}(z)$ and $H_{5/2}^{(1)}(z)$ 
through the elementary functions we get for $A_\mu$
\begin{equation}
\label{last}
\tilde A_\mu=\frac{8\pi G_6}{R^3}
\left(\frac{5c}{2k^2}+\frac{2(c-2ik)}{4k^2+6ick-3c^2}\right)q\tilde j_\mu
\end{equation}
 In the case of static configurations $p_0=0$ the first 
term in the last equation gives the conventional Coulomb law at large 
distances from a static source. The second term of (\ref{last}) 
results in corrections of order of $O(1/L^3)$ to Coulomb law at  
distances  $L\sim c^{-1}$ along the brane.

\section{Conclusion.}

We have considered models in which a brane universe $M^4$ is embedded 
into a space-time ${\cal M}^6$ with  $n=2$ warped infinite extra dimensions. 
The Kaluza-Klein field $A_\mu$ associated  to the $U(1)$ group of rotations 
of the normal bundle of $M^4$ is localized on the brane if the background 
metric (\ref{metric}) is such that the integral $N$ (\ref{integral}) is converging.
In the case of asymptotically anti-deSitter metric in  six-dimensional bulk
the fermion zero modes which possess nonzero rotation moment $q$ 
around the brane 
are localized on $M^4$ (see (\ref{rad})). 
They are charged with respect to $A_\mu$ and $A_\mu$ 
mediates Coulomb-like interaction between these zero modes. We have calculated 
the effective four-dimensional fine structure constant 
$\alpha$ (\ref{al}) in such a model. It is expressed through the 
inverse curvature radius $c$ of the bulk anti-deSitter space and the 
brane thickness $R$. The presence of infinite extra dimensions results in 
modification of the four-dimensional photon propagator and  
in power-law corrections to the Coulomb law at the distances of order of 
the inverse curvature radius $c$ of anti-deSitter space (\ref{last}).

\section{Acknowledgement.}

I am grateful to A.O.Barvinsky, V.F.Mukhanov, V.A.Rubakov, I.Sachs, 
and S.Solodukhin for fruitful discussions of the subject of the paper. This work 
was supported by SFB 375 der Deutschen Forschungsgemeinschaft.


\begin{thebibliography}{99}

\bibitem{appelquist} T.Appelquist, A.Chodos, P.G.O.Freund, eds.,
{\it Modern Kaluza-Klein theories},
Addison-Wesley, 1987

\bibitem{witten} E.Witten, 
``Fermion quantum numbers in Kaluza-Klein theory'', in \cite{appelquist}, 
p. 438

\bibitem{salam} S.Randjbar-Daemi, A.Salam, J.Strathdee, Nucl.\ Phys.\ 
{\bf B 214}, 491 (1983)

\bibitem{wetterich} C.Wetterich, Nucl.\ Phys.\ {\bf B 255}, 480 (1985)

\bibitem{rubakov} V.~A.~Rubakov and M.~E.~Shaposhnikov,
Phys.\ Lett.\ {\bf B125}, 136 (1983).

\bibitem{gibbons}
G.~W.~Gibbons and D.~L.~Wiltshire,
Nucl.\ Phys.\ {\bf B287}, 717 (1987).

\bibitem{dimopoulos}
N.~Arkani-Hamed, S.~Dimopoulos and G.~Dvali,
Phys.\ Rev.\ {\bf D 59}, 086004 (1999)
[hep-ph/9807344].

\bibitem{randall}
L.~Randall and R.~Sundrum,
Phys.\ Rev.\ Lett.\ {\bf 83}, 4690 (1999)
[hep-th/9906064].

\bibitem{hep}
T.~Han, J.~D.~Lykken and R.~Zhang,
Phys.\ Rev.\ {\bf D 59}, 105006 (1999)
[hep-ph/9811350].



\bibitem{astro}
S.~Cullen and M.~Perelstein,
Phys.\ Rev.\ Lett.\ {\bf 83}, 268 (1999)
[hep-ph/9903422]

\bibitem{cosmology} 
A.~Neronov,
Phys.\ Lett.\ {\bf B472}, 273 (2000)
[gr-qc/9911122]

\bibitem{ivo}
A.~Neronov and I.~Sachs,
``On metric perturbations in brane-world scenarios,''
hep-th/0011254.

\bibitem{fermion_hierarchy}
N.~Arkani-Hamed and M.~Schmaltz,
Phys.\ Rev.\ {\bf D 61}, 033005 (2000)
[hep-ph/9903417].


\bibitem{constant}
A.~Neronov,
``Brane world cosmological constant in the models with large extra  dimensions,''
gr-qc/0101060.

\bibitem{supersymmetry}
I.~Antoniadis, C.~Munoz and M.~Quiros,
Nucl.\ Phys.\ {\bf B397}, 515 (1993)
[hep-ph/9211309].

\bibitem{jackiw}
R.~Jackiw and C.~Rebbi,
Phys.\ Rev.\ {\bf D 13}, 3398 (1976).

\bibitem{dvali-shifman}
G.~Dvali and M.~Shifman,
Phys.\ Lett.\ {\bf B396}, 64 (1997)
[hep-th/9612128].

\bibitem{dubovsky}
S.~L.~Dubovsky, V.~A.~Rubakov and P.~G.~Tinyakov,
JHEP{\bf 0008}, 041 (2000)
[hep-ph/0007179].

\bibitem{oda}
I.~Oda,
Phys.\ Lett.\ {\bf B496}, 113 (2000)
[hep-th/0006203].

\bibitem{gababadze}
G.~Dvali, G.~Gabadadze and M.~Shifman,
Phys.\ Lett.\ {\bf B497}, 271 (2001)
[hep-th/0010071].



\bibitem{gregory}
R.~Gregory,
Phys.\ Rev.\ Lett.\ {\bf 84}, 2564 (2000)
[hep-th/9911015].

\bibitem{shaposhnikov}
T.~Gherghetta and M.~Shaposhnikov,
Phys.\ Rev.\ Lett.\ {\bf 85}, 240 (2000)
[hep-th/0004014].


\bibitem{shaposhnikov1}
T.~Gherghetta, E.~Roessl and M.~Shaposhnikov,
Phys.\ Lett.\ {\bf B491}, 353 (2000)
[hep-th/0006251].


\bibitem{giddings}
S.~B.~Giddings, E.~Katz and L.~Randall,
JHEP{\bf 0003}, 023 (2000)
[hep-th/0002091].




























\end{thebibliography}
\end{document}